\documentclass[conference]{IEEEtran}

\usepackage[frozencache]{minted}
\setminted{fontsize=\small}

\newcommand{\ident}[1]{$\texttt{#1}$}
\usepackage[usenames,dvipsnames]{xcolor}
\newcommand{\tlakw}[1]{{\color{OliveGreen}\textbf{#1}}}

\newcommand{\tlalabel}[1]{{\color{Cerulean}\textbf{#1}}}

\usepackage{todonotes}
\usepackage{xurl}
\usepackage{hyperref}

\begin{document}

\title{Combating Reentrancy Bugs on Sharded Blockchains}
\author{
\IEEEauthorblockN{Ognjen Mari{\'c}}
\IEEEauthorblockA{DFINITY Foundation\\
ognjen.maric@dfinity.org}
\and 
\IEEEauthorblockN{Robin K{\"u}nzler}
\IEEEauthorblockA{DFINITY Foundation\\
robin.kunzler@dfinity.org}
\and 
\IEEEauthorblockN{Lara Schmid}
\IEEEauthorblockA{DFINITY Foundation\\
lara.schmid@dfinity.org}
\and
\IEEEauthorblockN{Roman Kashitsyn}
\IEEEauthorblockA{Gensyn AI\\
roman.kashitsyn@gmail.com}

}
\maketitle

\begin{abstract}
Reentrancy is a well-known source of smart contract bugs on Ethereum, leading e.g.~to double-spending vulnerabilities in DeFi applications. 
But less is known about this problem in other blockchains, which can have significantly different execution models. Sharded blockchains in particular generally use an asynchronous messaging model that differs substantially from the synchronous and transactional model of Ethereum. 
We study the features of this model and its effect on reentrancy bugs on three examples: the Internet Computer (ICP) blockchain, NEAR Protocol, and MultiversX. We argue that this model, while useful for improving performance, also makes it easier to introduce reentrancy bugs. For example, reviews of the pre-production versions of some of the most critical ICP smart contracts found that 66\% (10/15) of the reviewed contracts — written by expert authors — contained reentrancy bugs of medium or high severity, with potential damages in tens of millions of dollars. We evaluate existing Ethereum programming techniques (in particular the effects-checks-interactions pattern, and locking) to prevent reentrancy bugs in the context of this new messaging model and identify some issues with them. We then present novel Rust and Motoko patterns that can be leveraged on ICP to solve these issues.
Finally, we demonstrate that the formal verification tool TLA+ can be used to find and eliminate such bugs in real world smart contracts on sharded blockchains.
\end{abstract}

\begin{IEEEkeywords}
security, reentrancy, formal verification, smart contracts, blockchain
\end{IEEEkeywords}

\section{Introduction}
A decentralized autonomous organization (DAO) called ``The DAO" was initiated in 2016 on the Ethereum blockchain as the world's first fully digital and decentralized investment fund. But it quickly came to a halt when an attacker exploited a reentrancy vulnerability and extracted \$50 million USD worth of Ether \cite{DAOattack}. Following The DAO incident, reentrancy bugs continued to surface in various Ethereum smart contracts. Uniswap and Lendf.Me lost \$25 million in 2020, CREAM \$18 million in 2021, and Fei \$80 million in 2022 due to similar vulnerabilities. 

Reentrancy bugs are a general class of concurrency bugs that long predates smart contracts. In Ethereum smart contracts, such bugs occur when a function (method) in a smart contract calls external contracts and causes an unintended interaction with its internal state. The prevalence of these issues stems from the inherent complexity in smart contract interactions, often involving multiple contracts with interdependent code. Detecting and mitigating such vulnerabilities poses a significant challenge due to the extensive number of potential interactions~\cite{EthereumReentrancyGuide}.

The persistence of this class of bugs has spawned a slurry of research papers on reentrancy \cite{SolidityPatterns,ReGuardFuzzer,ECF}, including work on security verification \cite{SOKSecuriyVerification,Fstar}.
The vast majority of this research has focused on Ethereum, which is unsurprising given its dominant market share. However, a number of more recent blockchains (e.g.,~\cite{ICPwhitepaper,NearWhitePaper,the_multiversx_team_multiversx_2019,wood_polkadot_nodate,hellings_cerberus_2020}) use a different, \emph{sharded} architecture in order to overcome the scalability bottleneck of the Ethereum architecture. A sharded blockchain is a collection of shards, where each shard can be seen as a subchain that maintains only a portion of the entire blockchain's state. Shards generally have disjoint validators, lowering the computational burden on each individual validator, as they only keep track of their relevant portion of the state.

\begin{figure*}[!t]
\begin{minted}[xleftmargin=20pt,linenos]{rust}
#[update]
async fn example_method(params: ...) -> ... {
    // first message handler m1
    let downstream_params = ...; // process params
    let downstream_result = call(downstream_contract, downstream_params).await;
    // second message handler (callback) m2
    let my_result = ...; // process downstream_result
    my_result
}
\end{minted}
\caption{Async/await for callback handling}
\label{fig:async-await}
\end{figure*}

To maximize throughput, shards normally operate asynchronously. Consequently, messaging between contracts on different shards is usually also asynchronous. This affects when and how reentrancy bugs can manifest themselves and how they can be addressed. 
Another difference compared to Ethereum is that many sharded blockchains use Web Assembly (Wasm) instead of EVM for their runtime and support Rust smart contracts, changing the space of possible solutions for reentrancy.
% Moreover, many sharded blockchains now use Web Assembly (Wasm) for their runtime, and support Rust smart contracts, changing the space of possible solutions compared to an EVM blockchain.

In this paper, we study reentrancy bugs in this new setting, focusing on blockchains where the interaction between contracts is done in a remote procedure call (RPC) style, which makes the interaction superficially resemble that of Ethereum contracts. To narrow down the solution space, we focus on blockchains that feature Wasm and Rust support. As examples of such blockchains, we study the Internet Computer blockchain (ICP)~\cite{ICPwhitepaper}, NEAR~\cite{NearWhitePaper}, and MultiversX~\cite{the_multiversx_team_multiversx_2019}. 

Our thesis is that this new setting makes reentrancy bugs even more likely to occur. For example, we reviewed the initial versions of the most important ICP smart contracts, such as the contracts governing the platform itself, DAO contracts governing the applications running on ICP, as well as contracts wrapping Bitcoin and other tokens. The reviews showed that 66\% (10/15) of these contracts — written by expert authors — contained reentrancy bugs of medium or high severity, with potential damages in tens of millions of dollars.
Thus, to profit from the increased performance of asynchronous messaging models while not compromising the security of smart contracts, it is of paramount importance to understand reentrancy bugs in those models and develop robust programming techniques to counteract such bugs. Furthermore, we need practical ways of ensuring that these programming techniques are applied correctly and actually prevent the bugs.

%blockchains But much less is known about such attacks on other blockchains, including the Internet Computer blockchain (ICP)\cite{ICPwhitepaper}. The ICP is particularly interesting since its asynchronous smart contract calls (aimed to boost scalability) also change how smart contracts interact with one another, which affects when and how reentrancy bugs can manifest themselves, and how they can be addressed. Lessons learned from studying the ICP would also translate to other blockchains that run calls asynchronously. 

%Like in ICP, cross-contract calls in NEAR\cite{NearWhitePaper} are asynchronous. In both cases the code that issues the cross-contract call registers a \textit{callback} method, but between issuing the call and executing the callback arbitrary other calls can be processed. Therefore, reentrancy bugs look very similar on both platforms. See NEAR's security docs on cross-contract calls\cite{NearCrossContractSecurityDocs} for an example of such an issue. 

%Finally, the Internet Computer and other blockchains support different programming languages, and in particular Rust, allowing us to explore different programming solutions to the reentrancy problem than those on Ethereum.

This paper makes the following contributions to the study of reentrancy:
\begin{enumerate}
    \item We describe how reentrancy bugs manifest themselves in the novel, asynchronous execution model of sharded blockchains (as opposed to the synchronous, transaction-based model of Ethereum) and the new challenges that this model poses.
    \item We identify new challenges for known programming techniques for preventing reentrancy, in particular the effects-checks-interactions pattern and locking~\cite{SolidityPatterns}.
    \item We describe how features of the Internet Computer can be leveraged to provide a robust locking technique using Rust and Motoko (a bespoke language for ICP~\cite{Motoko}).
    \item We demonstrate that formal verification, in particular model checking using the Temporal Logic of Actions (TLA+) toolkit, is effective in combating reentrancy bugs in the asynchronous messaging model. We show on a real-world case study how TLA+ can verify the correct application of the locking technique mentioned above. We also describe a general strategy for modeling Internet Computer smart contracts in TLA+.
\end{enumerate}

The paper is organized as follows. In section~\ref{sec:background}, we provide the background on the execution and messaging models of the studied blockchains. Section~\ref{sec:reentrancy} describes how reentrancy bugs manifest themselves in these models. Then, in section~\ref{sec:rust-techniques} we present existing programming techniques to prevent reentrancy bugs on Ethereum and discuss their shortcomings for sharded blockchains. We then provide concrete locking patterns in Rust and Motoko to prevent reentrancy bugs for smart contracts on the Internet Computer. In section~\ref{sec:tla}, we describe an approach to discover reentrancy bugs using TLA+ or verify their absence, including a general strategy for modeling Internet Computer smart contracts in TLA+. We evaluate the effectiveness of the patterns and TLA+-based verification in section~\ref{sec:evaluation}, and conclude in section~\ref{sec:conclusion}.

\section{Background}
\label{sec:background}

We start with some background on the execution and messaging models of the three blockchains that we study in this paper: the Internet Computer, the NEAR protocol, and MultiversX.

\subsection{Internet Computer (ICP)}
\label{sec:bg-ic}

Smart contracts on ICP are written using WebAssembly (Wasm), or using any language that can compile down to Wasm, such as Rust or Motoko. In addition to its Wasm heap and stack, an ICP smart contract gets access to ICP's system API using Wasm module imports. 

The most important aspect of the system API for our purposes is related to contract calls. Logically, ICP allows a smart contract to expose \emph{methods} that other contracts (and users) can call. A method then provides a response to the caller, which the caller can process. Under the hood, a call is performed using two \emph{messages}: a \emph{request} from the caller to the callee, and a \emph{response} from the callee to the caller. Contracts process these messages using \emph{message handlers}, which are just Wasm functions. Such functions can be marked as public, making them method entry points. While processing a message (either a request or a response), a handler can issue a response, or it can issue further downstream calls. For each such downstream call, the handler designates a Wasm function from the contract code as a \emph{callback function}, which the system will use as the message handler for the response (the function needn't be public). Thus, a method call can be broken up into multiple message handlers. Crucially, whenever some handler issues a downstream request, the execution of the upstream call is effectively suspended until the response arrives, but the contract is allowed to process other messages (both other requests and responses).

Moreover, higher-level languages (such as Rust) often come with syntactic sugar in the form of async/await syntax, which can be leveraged to take care of the callback handling for the user. For example, the Internet Computer Rust libraries support writing code as in Figure~\ref{fig:async-await}. This code is compiled into two message handlers, the first of which processes the parameters and then makes the downstream call. The second handler processes the result of the downstream call and returns the result to the caller.

This messaging model has the following properties that are important for the topic of reentrancy:
\begin{enumerate}
    \item Only a single message handler (handling a single message) is executed at a time per smart contract. So message execution is sequential, and never parallel.
    \item Multiple messages, e.g., from different calls, can interleave and have no reliable execution ordering.
    \item If a message handler traps / panics, its state changes are reverted.
    \item Every request will eventually receive a response, though the response may be a system-generated error response.
\end{enumerate}
%For further detail refer to the Internet Computer Interface Specification\cite{InterfaceSpec}, in particular the sections on ``Abstract Behaviour" and ``Ordering Guarantees".

\begin{figure*}[!t]
\begin{minted}[xleftmargin=20pt,linenos]{rust}
#[update]
async fn notify_minter() -> Result<(), String> {
    let owner = caller(); 
    let utxos = fetch_new_utxos(owner).await?;
    mint_ckbtc_on_ledger(owner, utxos).await?;
    minter_controlled_utxos.update(owner, utxos);
    Ok(())
}
\end{minted}
\caption{A simplified version of the real-world minter notification method}
\label{fig:vulnerable-notify-minter}
\end{figure*}

Consider Figure~\ref{fig:async-await} again, and suppose that two calls $A$ and $B$ are made to \texttt{example\_method} roughly at the same time. As discussed, for each call two message handlers are executed. For call $A$, we denote the first message handler (line~3 to 6) as $m_{A,1}$ and the callback (line~5 to 9) as $m_{A,2}$. Analogously, the handlers for call $B$ are denoted as $m_{B,1}$ and $m_{B,2}$. Let's consider possible message execution orderings in this example, according to the messaging properties above. Probably the most intuitive ordering is $m_{A,1}, m_{A,2}, m_{B,1}, m_{B,2}$, i.e.~each callback is executed immediately after the first message. But as there is no reliable message ordering (property~2), for example, the execution ordering $m_{A,1}, m_{B,1}, m_{A,2}, m_{B,2}$ is also possible. This second ordering, where both callbacks are executed only after processing the first messages, can be highly beneficial as it can greatly increase throughput. For example, $m_{B, 1}$ can be executed in parallel with the downstream contract's handler that's handling A's downstream call. However, the ordering can be unintuitive and unexpected by developers, leding to subtle reentrancy vulnerabilities, as we describe later.

\subsection{NEAR Protocol}
\label{sec:bg-near}

The execution and messaging models of NEAR are very similar to those of ICP, as demonstrated by the following properties.
Smart contracts in NEAR are written in Wasm, or any language that compiles to Wasm, with official SDKs for Rust and Typescript. 
Cross-contract calls in NEAR are asynchronous and the code that issues the cross-contract call registers a \textit{callback} method.
Between issuing the call and executing the callback, arbitrary other calls can be processed. 
Furthermore, every call is guaranteed to eventually receive a response, and a message handler that traps reverts its state. 

Despite these similarities, there are two notable differences in NEAR compared to ICP:
\begin{itemize}
    \item There is no support for async/await Rust syntax; callbacks must be explicitly constructed.
    \item Callbacks must be publicly accessible, so they have to do their own authorization checks.
\end{itemize}

Due to the models being so similar, reentrancy bugs look very similar on both platforms.

%Like in ICP, smart contracts in NEAR are written in Wasm, or any language that compiles to Wasm, with official SDKs for Rust and Typescript. Again like in ICP, cross-contract calls in NEAR are asynchronous. In both cases the code that issues the cross-contract call registers a \textit{callback} method, but between issuing the call and executing the callback arbitrary other calls can be processed. Furthermore, just like in ICP, every call is guaranteed to eventually receive a response, and a message handler that traps reverts its state. The model is thus very similar to the one of the ICP, with the following differences:
%\begin{itemize}
%    \item There is no support for async/await Rust syntax; callbacks must be explicitly constructed.
%    \item Callbacks must be publicly accessible, so they have to do their own authorization checks.
%\end{itemize}
%Therefore, reentrancy bugs look very similar on both platforms.
%See NEAR's security docs on cross-contract calls\cite{NearCrossContractSecurityDocs} for an example of such an issue. 

\subsection{MultiversX}
\label{sec:bg-multiversx}

The execution and messaging models of MultiversX are almost identical to those of NEAR. Wasm and Rust are supported, and callbacks must be constructed explicitly, with explicit authorization checks. The main difference is that MultiversX also supports an additional, synchronous call model, which works only for calling contracts on the same shard.

\section{Reentrancy attacks} 
\label{sec:reentrancy}
\iffalse
(general and in the IC model -2 pages)

- use ckBTC as the running example? this is currently assumed by the TLA part; it's assumed that it's described what the "ckBTC minter" and "ckBTC ledger" are, and that the attack is described.
- note: the ckBTC example is actually kinda cool because it shows how we can process two user calls in parallel, something that doesn't exist in Ethereum, and also that we can have new kinds of reentrancy bugs (due to such parallel calls)
\fi

\begin{figure*}[t]
\begin{minted}[xleftmargin=20pt,linenos]{rust}
#[update]
async fn notify_minter() -> Result<(), String> {
    assert!(!is_locked(), "Already processing a call");
    lock();
    let owner = caller();
    let utxos = fetch_new_utxos(owner).await?;
    mint_ckbtc_on_ledger(owner, utxos).await?;
    minter_controlled_utxos.update(owner, utxos);
    unlock();
    Ok(())
}
\end{minted}
\caption{Basic mutex-based locking pattern}
\label{fig:basic-locking}
\end{figure*}

We now describe new flavors of reentrancy issues that can arise due to the properties of the messaging models described in section~\ref{sec:background}. To illustrate this, we give a simplified example, inspired by an actual vulnerability we found in the ckBTC contracts. \emph{ckBTC}~\cite{ckBTC} is a token on the Internet Computer that is securely backed 1:1 by Bitcoin\cite{SatoshiWhitepaper} (BTC), allowing faster and cheaper transactions compared to the Bitcoin network. The \emph{ckBTC ledger} smart contract manages the users' balances of ckBTC. To obtain ckBTC, users deposit BTC on the Bitcion network to an account owned by the \emph{ckBTC minter}. The minter uses ICP's chain-key threshold ECDSA signatures \cite{tECDSA} to own Bitcoin. Once the user deposits BTC, it notifies the minter contract, which converts the deposited BTC to ckBTC.

The simplified version of the minter's notification method is shown in Figure~\ref{fig:vulnerable-notify-minter}. On line~4, the minter fetches the caller's deposited UTXOs~\cite{UTXOPaper} from the Bitcoin network, with the help of a special BTC smart contract that integrates ICP with Bitcoin. Only UTXOs that have not yet been processed and stored in the internal data structure \texttt{minter\_controlled\_utxos} are considered as new and are used to mint ckBTC. The minter then instructs the ledger on line~5 to mint the corresponding amount of ckBTC for the caller. Finally, the newly minted UTXOs are accounted for in the minter's internal data structure on line~6. 

An important security goal of the minter is to avoid double-minting by making sure that newly deposited UTXOs can be used at most once to mint ckBTC. At first sight, \texttt{notify\_minter} seems to achieve that, because the new UTXOs are accounted for immediately after the ledger is instructed to mint the ckBTC, and the next invocation of the method will no longer consider these UTXOs as new. 

However, there is a subtle reentrancy issue that allows double minting. Consider two parallel calls $A$ and $B$ again. For simplicity, we ignore that line~4 also calls out to another contract, as that is not important for our example. With this simplification, just as in the example of section~\ref{sec:bg-ic}, there are again message handlers $m_{A,1}, m_{A,2}$ and $m_{B,1}, m_{B,2}$ corresponding to the two calls, respectively, where the second handlers are the callback executions that update \texttt{minter\_controlled\_utxos} on line~6. Now suppose the handlers are executed in the ordering $m_{A,1}, m_{B,1}, m_{A,2}, m_{B,2}$. Then, $m_{A,1}$ and $m_{B,1}$ both fetch the caller's UTXOs, and both consider the newly deposited UTXOs as new, resulting in double-minting the given amount of ckBTC on the ledger in line~5. Only in the callback executions $m_{A,2}, m_{B,2}$, when it is already too late, is the internal data structure updated to prevent later calls from considering the same UTXOs as new. 

This interleaving could also arise in NEAR and MultiversX, the code would just look differently as these chains do not support the async/await syntax.
However, this interleaving \emph{could not occur} in an Ethereum smart contract, as the entire call including the mint operation on the ckBTC ledger would be executed atomically, with no interleaved calls. If the ledger is trusted -- which it would normally be, as it is developed and controlled by the same entity as the minter contract -- it will also not call back into the minter contract, so the code above is actually safe. But if the ledger were not trusted, the analogous minter contract on Ethereum would also have a similar double-minting issue if the ledger maliciously called back into \texttt{notify\_minter}~\cite{EthereumReentrancyGuide}. We will next see what the existing approaches to counteracting such bugs on Ethereum are, and why these approaches cannot be directly translated to the asynchronous models from section~\ref{sec:background}. 

\section{Programming techniques for combating reentrancy attacks} 
\label{sec:rust-techniques}

Reentrancy bugs can be avoided by using safe coding patterns. Demeyer et al.~\cite{SolidityPatterns} identify two main such patterns for Solidity and the Ethereum blockchain: \textit{checks-effects-interactions} and \textit{mutexes}.

The checks-effects-interactions pattern modifies a contract's internal data structures before interacting with external contracts. Consider again the \texttt{notify\_minter} example above, but now consider the case where the ledger is potentially malicious and can call back into the minter contract. As mentioned earlier, this could cause the double-minting issue above even on Ethereum. To counteract that, the checks-effects-interactions pattern would swap the order in the code above to put the \textit{effect} of updating the internal data structure (line~6) before the \textit{interaction} of calling the ledger (line~5). However, this fix does not trivially work in the asynchronous models from section~\ref{sec:background}: it could happen that the UTXOs would be accounted for (first message handler) but then the mint call fails and no ckBTC are minted. Unlike on Ethereum, the effects of updating the internal data structures would then not be rolled back and
%(which by property 4 can always happen). 
these internal data structures would be left in an inconsistent state.
%but it can happen on any of the other blockchains covered in Section~\ref{sec:background}, even if the code would look differently due to a lack of async/await syntax.

The second pattern, mutex, prevents concurrent calls by obtaining a mutual exclusion lock on the minter contract. We can apply it to the \texttt{notify\_minter} example as shown in Figure~\ref{fig:basic-locking}. There, the added locking functions check/set/unset a simple boolean flag in the smart contract's state. The same pattern would be applied to any state-changing function on the minter contract, ensuring that no interleaved calls are possible, and thus preventing reentrancy attacks. However, there are two problems with directly applying this pattern to the blockchains covered in section~\ref{sec:background}:
\begin{enumerate}
    \item This type of locking can be brittle. In all of the blockchains covered in section~\ref{sec:background}, if the callback traps (e.g.~due to a bug), state changes are not persisted and thus the contract is left in a locked and broken state.
    \item Such contract-wide locking as above is normally too aggressive as it negates the performance benefits of interleavings that were mentioned in section~\ref{sec:background}. Finer-grained locking is often needed to achieve both security and performance.
\end{enumerate}

In the following, we describe locking patterns for Internet Computer contracts to avoid the issues above. We first provide a detailed Rust pattern which is inspired by Alexandrescu~\cite{AlexandrescuLockingPatterns}, and then give a high-level description of a pattern in Motoko~\cite{Motoko}, a language specifically designed for ICP. The code in Figure~\ref{fig:improved-locking} shows how to implement a lock per caller (\texttt{CallerGuard}) with a \texttt{Drop} implementation that automatically releases the lock when the struct goes out of scope. As can be seen in the modified \texttt{notify\_minter} method in Figure~\ref{fig:improved-locking}, the lock is very easy to use.

\begin{figure*}
\begin{minted}[xleftmargin=20pt,linenos]{rust}
pub struct State {
    // A "principal" is an ICP user identifier
    pending_requests: BTreeSet<Principal>,
}
thread_local! {
    static STATE: RefCell<State> = RefCell::new(
        State{pending_requests: BTreeSet::new()}
    );
}
pub struct CallerGuard {
    principal: Principal,
}
impl CallerGuard {
    pub fn new(principal: Principal) -> Result<Self, String> {
        STATE.with(|state| {
            let pending_requests = &mut state.borrow_mut().pending_requests;
            if pending_requests.contains(&principal){
                return Err(format!("Already processing a request for principal {}", 
                                    &principal));
            }
            pending_requests.insert(principal);
            Ok(Self { principal })
        })
    }
}
impl Drop for CallerGuard {
    fn drop(&mut self) {
        STATE.with(|state| {
            state.borrow_mut().pending_requests.remove(&self.principal);
        })
    }
}
#[update]
async fn notify_minter() -> Result<()), String> {
    let owner = caller();
    // Using `?`, return an error immediately if there is already a call in progress for 
    // `caller`.
    // Note that `let _ = CallerGuard::new(caller)?`, wouldn't work as it will drop 
    // the guard immediately and locking would not be effective.
    let _guard = CallerGuard::new(owner)?;
    let utxos = fetch_new_utxos(owner).await?;
    mint_ckbtc_on_ledger(owner, utxos).await?;
    minter_controlled_utxos.update(owner, utxos);
    Ok(())
} // Here the guard goes out of scope and is dropped
\end{minted} 
\caption{Improved locking pattern}
\label{fig:improved-locking}

\end{figure*}

The pattern from Figure~\ref{fig:improved-locking} can be adapted to the appropriate locking granularity, e.g.~to implement a global (contract-wide) lock, to acquire a lock on any subset of the smart contract's state, or for limiting the number of callers that are allowed to execute a given method at the same time. 

The pattern also avoids the other issue mentioned earlier, namely that the smart contract could end up in a locked state if the callback traps. It avoids this problem by using ICP's system API and Rust libraries features that can call cleanup code on traps in order to make sure that any local variables are still dropped, in this case dropping the lock. Note that this cleanup code is not visible in the above example, because it is part of ICP's smart contract Rust library that automatically registers the cleanup. While one can use finer-grained locks and release them in callbacks in both NEAR and MultiversX as well, the pattern presented above is unfortunately not applicable to them, as there is no functionality that allows dropping the lock on trap. In case of MultiversX, it would be possible to pass the lock to the callback where it would get dropped on successful execution of the callback, but still the lock would not be released on a trap.

In Motoko, an analogous locking behavior can be achieved by using \texttt{try} / \texttt{finally}~\cite{MotokoTryFinally}. If errors or traps happen in the \texttt{try} or \texttt{catch} blocks, the \texttt{finally} clause will still execute and can safely release locks. This employs the same cleanup functionality mentioned above.

\section{TLA+ for reentrancy attacks} 
\label{sec:tla}

Applications of the locking technique should be evaluated to confirm that they actually correctly prevent reentrancy bugs.
As such bugs occur due to unexpected interleavings of smart contract calls, and as the number of possible interleavings is normally huge, it is difficult to establish the correctness of a smart contract using simple testing, even if test coverage is high. Techniques like fuzzing are also difficult to apply, since they would require controlling the message handler scheduling, which none of our surveyed blockchains support. An alternative approach is to apply formal verification tools to systematically (within some bounds) examine all possible schedules, and find such bugs (including any intentional exploits) or ensure their absence.
In this section, we show how to use TLA+ for this purpose, on the example of the ckBTC minter contract.

TLA+~\cite{lamport2002specifying} is a formal language for specifying and verifying complex systems. Additionally, the language is accompanied by a set of tools for formal verification. The tools include TLC~\cite{tlc} (an explicit state model checker), Apalache~\cite{konnov2019tla} (an SMT-based bounded model checker), and TLAPS~\cite{cousineau2012tla} (an interactive proof assistant for TLA+).

TLA+ models systems as transition systems, also called state machines. While the logical formulas of TLA+ can have many different shapes, in practice, systems are described in TLA+ by providing the following four elements:
\begin{enumerate}
    \item A list of \emph{variables} that the system state consists of.
    \item An \emph{initial predicate} \ident{Init}, which specifies the allowed initial states by specifying the constraints on variables.
    \item A \emph{transition predicate} \ident{Next}, which specifies how the state of the system evolves. It is a predicate over two states, the current and the successor state, and the system is allowed to move from a state $s$ to a state $s'$ whenever \ident{Next}$(s, s')$ holds.
    \item The system can optionally specify some \emph{fairness conditions}, that are useful when specifying liveness properties, as we will see later.
\end{enumerate}

As such, TLA+ is very general and not in any way geared towards any particular messaging model for smart contracts, or smart contracts in general. However, the asynchronous model in particular can be modeled very easily. Transition system interleavings can directly model message handler interleavings. Thus, TLA+ can be used to model contracts in any of the blockchains from section~\ref{sec:background}, for the purpose of eliminating reentrancy bugs. Moreover, the TLA+ toolkit also provides PlusCal~\cite{lamport2009pluscal}, which is essentially syntax sugar on top of TLA+, and which we found to be well suited for modeling Internet Computer smart contracts that use the async/await syntax. We next describe a general strategy for this modeling.

\subsection{Using PlusCal to model async/await}

\begin{figure*}
\begin{minted}[xleftmargin=20pt,linenos,escapeinside=||]{text}
|\tlakw{variables}| 
    minter_btc_requests = <<>>; minter_btc_responses = {};
    minter_ledger_requests = <<>>; minter_ledger_responses = {};
    minter_controlled_utxos = {}; locks = {};

|\tlakw{process}| ( Notify_Minter \in NOTIFY_MINTER_PROCESS_IDS) 
    |\tlakw{variable}| owner \in USERS, utxos = {};
{
|\tlalabel{Notify\_Minter\_Start}|:
    |\tlakw{await}|(owner \notin locks); |\label{line:await}|
    locks := locks \union {owner};
    make_fetch_utxo_request(|\tlakw{self}|, owner);
|\tlalabel{Notify\_Minter\_Receive\_Utxos}|:
    |\tlakw{with}|(response \in { r \in minter_btc_responses: Caller(r) = |\tlakw{self}|}) {
      minter_btc_responses := minter_btc_responses \ {response};
      |\tlakw{if}|(Is_Ok(response.status)) { 
        utxos := Get_Utxos(response);
        make_minting_request(|\tlakw{self}|, owner, utxos);
      } |\tlakw{else}| {
        locks := locks \ {owner};
        goto Done;
      }
    }; 
|\tlalabel{Notify\_Minter\_Mark\_Minted}|:
    |\tlakw{with}|(response \in { r \in minter_ledger_responses: Caller(r) = |\tlakw{self}|}) {
        minter_ledger_responses := minter_ledger_responses \ {response};
        |\tlakw{if}|(Is_Ok(response.status)) {
            update_minter_controlled_utxos(owner, utxos);
        };
    };
    locks := locks \ {owner};
    goto Done;
};
    
\end{minted}
    \caption{PlusCal/TLA model of \texttt{notify\_minter} method.}
    \label{fig:notify-pluscal}
\end{figure*}

In PlusCal, the transition system is described as a collection of \emph{processes}, written in an imperative style. Each process can have multiple concurrently executing \emph{instances}. In addition to global system variables, which are shared among the processes, each process can define its own local variables, and each process instance then has exclusive access to its own instance of those local variables. The model author can define how the variables are initialized. Processes contain one or more blocks, which are delimited using labels in the source code. Each block is executed atomically, but the execution of the different process instances can be interleaved by interleaving the execution of their blocks. 

Under the hood, PlusCal models get translated into a regular TLA+ transition system. The system's initial predicate constrains each variable to equal its initial value as defined in PlusCal. The system's transition predicate is defined as a disjunction of predicates, each of which describes how and when the execution of a single block of a single process instance can change the instance-local and global variables.

The insight here is that the execution model of the Internet Computer maps well onto the shape of PlusCal models, leading to the following general strategy of modeling Internet Computer smart contracts using PlusCal.

\paragraph{Modeling messages and contract-global state}
    The model uses global system variables for:
    \begin{enumerate}
        \item buffers with in-flight messages \emph{to} the contract;
        \item buffers with in-flight messages \emph{from} the contract; and
        \item the contract’s global state (i.e., its Wasm heap).
    \end{enumerate} 
    ICP orders requests delivered between a pair of smart contracts. We thus model the in-flight requests from one contract to another by a buffer variable that contains a queue and create one such buffer variable for each pair of contracts where the first contract calls the second. For example, Figure~\ref{fig:notify-pluscal} shows the PlusCal model of the \texttt{notify\_minter} method. There, we have one queue for the requests to the BTC canister that integrates with the Bitcoin network to provide the latest Bitcoin state, and another queue for requests to the ckBTC ledger. For responses, there is no ordering guarantee. The buffer variables that model responses thus hold an unordered collection, i.e., a set of responses. The variables are initialized such that the buffers are empty. An empty sequence, i.e., queue, is written as \texttt{<<>>} in TLA+. The requests (resp. responses) are then consumed (resp. produced) by another part of the model that models the BTC and ckBTC ledger canisters, which we omit here for brevity. As in the previous examples, we also omit the definitions of some auxiliary PlusCal functions (such as those for putting the requests in the buffers, or for updating the variable holding the minter controlled UTXOs). The two additional global variables in Line 4 of Figure~\ref{fig:notify-pluscal} model the ckBTC minter heap, storing the internal UTXO bookkeeping data and the locks. The initial values of these variables correspond to the state of the Wasm memory of a freshly installed smart contract, whatever that is for the particular contract (in this case, empty sets).

\paragraph{Modeling contract methods and local variables} We create one PlusCal process for each method of the analyzed smart contract. Each instance of the process will correspond to a single method execution. We can control how many executions are analyzed by setting the size of the instance (e.g, \texttt{NOTIFY\_MINTER\_PROCESS\_IDS} in Figure~\ref{fig:notify-pluscal}) when doing the verification (see section~\ref{sec:evaluation}). We rely on PlusCal process-local variables to model the method’s local variables. The variables are initialized on a case-by-case basis, and can also be chosen non-deterministically (e.g., an arbitrary user calling the method in reality is modeled by picking the owner non-deterministically in Figure~\ref{fig:notify-pluscal}).

\paragraph{Modeling atomic message handlers} We create one PlusCal label for each message handler in the method. E.g., we have the three labels start, receive UTXOs, and a mark minted for \texttt{notify\_minter} in Figure~\ref{fig:notify-pluscal}. Recall that blocks, delimited by labels, are executed atomically -- exactly like message handlers in ICP's execution model. The label body describes how the handler processes a message. We use PlusCal control flow constructs (such as if/else, goto, or just simple order of labels) to ensure that the model executes the different message handlers in the same order as the actual smart contract method does. We desugar the async/await syntax by explicitly removing a response from the incoming buffer at the start of each message handler (except the first), and enqueueing a new request at the end of each message handler (except the last). We use the process identifier (provided by the \texttt{self} keyword) to distinguish messages sent from concurrent invocations of the same method. We use the \texttt{with} keyword to non-deterministically pick responses satisfying a condition (the transition is blocked until the condition can be satisfied). Similarly, we use \texttt{await} to block transitions on a certain condition (e.g., we block the start of the \texttt{notify\_minter} method in line~\ref{line:await} as long as the owner is locked). We manually release locks in all cases.

\subsection{Defining properties}
\label{tla-properties}

In addition to the model of the smart contract, one must also specify the contract's desired properties. This is done in "pure" TLA+, i.e., without using PlusCal. We thus expect this part of the process to also apply to NEAR or MultiversX smart contracts, even if they cannot benefit from PlusCal. On the example of ckBTC (see the full TLA+ model~\cite{ckBTCModel} for more details), the main property is that the sum of the amounts of all UTXOs that are controlled by the ckBTC minter must be equal or higher than the total supply of ckBTC (as minted on the ckBTC ledger). This property protects the system from malicious users, by ensuring that they cannot exploit a reentrancy bug in order to double-spend their deposited BTC. We have found a violation of this property in an early version of the ckBTC contract. Interestingly, we have also observed reentrancy bugs that would be detrimental to the user. Namely, the early version of the ckBTC  contract also had a bug that could cause the user to lose their money in the process of redeeming BTC for ckBTC. The property that was violated by the bug was that, ignoring fees, if the system quiets down by no longer adding deposits or redeem requests, we want the total supply of ckBTC to eventually rise to equal the amount of deposited BTC. This way, we can guarantee that no BTC is lost in the conversion process. This is a liveness property, and it is conditioned on certain fairness conditions, which prohibit the model of the system from idling when there is work to be done (which doesn't happen in practice), but also require the users to sufficiently often update their ckBTC balances by calling the \ident{notify\_minter} method on the minter contract.

\section{Evaluation}
\label{sec:evaluation}

Having described both our programming techniques and our TLA+-based modeling and verification approach, we now evaluate the effectiveness of both. 

On the empirical side, we note that 8 of the evaluated Rust ICP contract suites, including 7 out of 9 previously vulnerable ones, are now using the proposed locking solution in production. Most of these suites include platform-critical contracts. The resulting smart contracts have no known bugs, while the performance impact of the locking patterns is insignificant. First, the pure instructions overhead is small: for the \texttt{notify\_minter} method, the locking code takes 8192 instructions in tests in small settings (few users and UTXOs), which currently costs around $4 \cdot 10^{-8}$ USD and represents 2.92\% of the method's total instructions on the happy path. Second, even though the locking pattern does in theory decrease the possible parallelism in processing a single user's requests (since it prohibits a user from running multiple operations), this has not caused any known performance problems in practice. The reason is that typical end-users rarely perform parallel operations. 

Furthermore, as discussed earlier, we used the TLA+ models to strengthen the confidence in the correctness of the smart contracts which use the new locking techniques.
Once the model and the properties are defined (as described in section~\ref{sec:tla}), the TLA+ toolkit can analyze whether the properties are satisfied by the model. For the analysis, we opted to use the TLC model checker. The model checker is able to perform the analysis fully automatically, with no further human input needed. This is in contrast to TLAPS, which requires significant manual interaction, and Apalache, which in practice requires the model author to formulate auxiliary inductive invariants, and also cannot analyze liveness properties. 

As with any explicit state model checker, the downside of using TLC is that it can only cope with a finite system state. This requires stating explicit bounds on all of the system's parameters. For example, TLC requires a bound on the number of concurrent calls to \texttt{notify\_minter}, even if in reality there is no hard bound. Similarly, bounds are required for the number of concurrent calls to the other methods, the total amount of BTC in circulation, the number of users considered, and so on. Furthermore, such bounds also have to be made very small, otherwise TLC quickly runs out of resources (such as memory or time), as any increase of the bounds tends to lead to an exponential blowup of the system space. For the ckBTC example, we have been able to verify all properties discussed in section~\ref{tla-properties} for a model~\cite{ckBTCModel} with up to 2 parallel invocations of the minter notification call, 2 BTC withdrawal calls (not presented in this paper), 2 users, and a BTC supply of 3 Satoshi. The model checking took around a day and further increases of bounds presumably would lead to further blow up of checking time. While these bounds may seem impractically small, we note that previous empirical research in different areas has found that the vast majority of bugs can be recreated with very small bounds. For example, Lu et al.~\cite{lu_learning_2008} found that 96\% of the studied concurrency bugs require only 2 threads to be triggered, and Yuan et al.~\cite{yuan_simple_2014} find that 98\% of the examined bugs in distributed systems can be triggered with 3 or fewer nodes. In our experience, the analysis with small bounds proved sufficient in practice. The analysis detects problems whenever the protective mechanisms are removed from the model. Furthermore, while we have found previously undiscovered bugs using TLC, we are yet to manually find a reentrancy bug that was missed in the TLC analysis due to the bounds being too small. 

Finally, we deem the TLA+ verification process to be fairly effective in terms of effort. A first model of the ckBTC minter contract took around 3 weeks to produce, which is comparable to the time needed for a manual security review. The scope of a manual review would be significantly larger, covering all security aspects rather than just reentrancy, but in our experience it is a reasonable investment given the prevalence of reentrancy bugs and their large impact, especially for DeFi applications where stakes are high.

\section{Conclusion}
\label{sec:conclusion}
In this paper, we studied the reentrancy problem in sharded blockchains, notably ICP, NEAR, and MultiversX.
In contrast to blockchains like Ethereum, these blockchains have asynchronous messaging models. Such models can exhibit reentrancy bugs in more scenarios than the Ethereum's synchronous model. They also make it more challenging to use existing techniques for preventing reentrancy bugs, including locking techniques.
First, 
%it is an additional challenge to deal with traps 
when locks are acquired in methods that span multiple messages, partial state rollbacks due to traps might cause locks to not be released.
Second, locking has to be balanced with performance: a major advantage of interleaving messages is that this increases performance, which is lost again if locking is done too aggressively.

For the Internet Computer, we provided a concrete locking pattern that prevents reentrancy bugs despite these challenges.
In future work, it would be interesting to develop similar patterns for NEAR and MultiversX.
Of course such patterns are only helpful if they are readily used in practice. Therefore, another relevant piece of future work is to make such patterns available in libraries so that they can be used and improved by many developers.

Since reentrancy problems are hard to find, we also demonstrated how one can leverage the formal verification tool TLA+ to help with this, using the ckBTC minter contract as a real-world case study.
We presented how the smart contracts on the Internet Computer written using async/await syntax can be conveniently translated into PlusCal, which in turn can be input to the TLA+ tool. We believe that TLA+ can also be effectively applied to NEAR and MultiversX where the message and execution models are similar; there, the modeling should be simpler, as there is no need to "desugar" the async/await syntax.
%could also define state as messages and state of smart contracts and each of calls and their callbacks as process. The way how to identify the callbacks would just be different
%(rather than find async/await in code, would find explicitly callbacks)
Another interesting question to further investigate is how models can be kept up-to-date with code, especially for mutable smart contracts that are developing over time. In particular, it would be interesting to study whether parts of the translation from code to models can be automated or whether discrepancies between them could be automatically detected, e.g., with testing.

\iffalse
Future work
- Provide locking libraries in Rust that are easy to use so people don't have to implement locks themselves
- TLA 
\fi

\newpage

%%%%%%%%%%%%%%%%%%%%%%%%%

%AUTHOR: comment out if using thebibliography
%\theendnotes

%AUTHOR: please read ledgerbib.bst usage notes by opening it in a text editor. We have modified it to include the use of the @misc item type for the proper formatting of online sources.

\bibliographystyle{IEEEtran}
\bibliography{references}

\begin{thebibliography}{10}
\providecommand{\url}[1]{#1}
\csname url@samestyle\endcsname
\providecommand{\newblock}{\relax}
\providecommand{\bibinfo}[2]{#2}
\providecommand{\BIBentrySTDinterwordspacing}{\spaceskip=0pt\relax}
\providecommand{\BIBentryALTinterwordstretchfactor}{4}
\providecommand{\BIBentryALTinterwordspacing}{\spaceskip=\fontdimen2\font plus
\BIBentryALTinterwordstretchfactor\fontdimen3\font minus
  \fontdimen4\font\relax}
\providecommand{\BIBforeignlanguage}[2]{{%
\expandafter\ifx\csname l@#1\endcsname\relax
\typeout{** WARNING: IEEEtran.bst: No hyphenation pattern has been}%
\typeout{** loaded for the language `#1'. Using the pattern for}%
\typeout{** the default language instead.}%
\else
\language=\csname l@#1\endcsname
\fi
#2}}
\providecommand{\BIBdecl}{\relax}
\BIBdecl

\bibitem{DAOattack}
\BIBentryALTinterwordspacing
N.~Atzei, M.~Bartoletti, and T.~Cimoli, ``A survey of attacks on ethereum smart
  contracts sok,'' in \emph{Proceedings of the 6th International Conference on
  Principles of Security and Trust - Volume 10204}.\hskip 1em plus 0.5em minus
  0.4em\relax Berlin, Heidelberg: Springer-Verlag, 2017, p. 164–186.
  [Online]. Available: \url{https://doi.org/10.1007/978-3-662-54455-6_8}
\BIBentrySTDinterwordspacing

\bibitem{EthereumReentrancyGuide}
D.~Muhs, ``Smart contract security field guide - reentrancy,'' (accessed April
  15 2024) \url{https://scsfg.io/hackers/reentrancy/}, 2023.

\bibitem{SolidityPatterns}
S.~Demeyer, H.~Rocha, and D.~Verheijke, ``Refactoring solidity smart contracts
  to protect against reentrancy exploits,'' in \emph{Leveraging Applications of
  Formal Methods, Verification and Validation. Software Engineering},
  T.~Margaria and B.~Steffen, Eds.\hskip 1em plus 0.5em minus 0.4em\relax Cham:
  Springer Nature Switzerland, 2022, pp. 324--344.

\bibitem{ReGuardFuzzer}
\BIBentryALTinterwordspacing
C.~Liu, H.~Liu, Z.~Cao, Z.~Chen, B.~Chen, and B.~Roscoe, ``Reguard: Finding
  reentrancy bugs in smart contracts,'' in \emph{Proceedings of the 40th
  International Conference on Software Engineering: Companion Proceeedings},
  ser. ICSE '18.\hskip 1em plus 0.5em minus 0.4em\relax New York, NY, USA:
  Association for Computing Machinery, 2018, p. 65–68. [Online]. Available:
  \url{https://doi.org/10.1145/3183440.3183495}
\BIBentrySTDinterwordspacing

\bibitem{ECF}
\BIBentryALTinterwordspacing
S.~Grossman, I.~Abraham, G.~Golan-Gueta, Y.~Michalevsky, N.~Rinetzky, M.~Sagiv,
  and Y.~Zohar, ``Online detection of effectively callback free objects with
  applications to smart contracts,'' \emph{Proc. ACM Program. Lang.}, vol.~2,
  no. POPL, dec 2017. [Online]. Available:
  \url{https://doi.org/10.1145/3158136}
\BIBentrySTDinterwordspacing

\bibitem{SOKSecuriyVerification}
J.~Liu and Z.~Liu, ``A survey on security verification of blockchain smart
  contracts,'' \emph{IEEE Access}, vol.~7, pp. 77\,894--77\,904, 2019.

\bibitem{Fstar}
\BIBentryALTinterwordspacing
K.~Bhargavan, A.~Delignat-Lavaud, C.~Fournet, A.~Gollamudi, G.~Gonthier,
  N.~Kobeissi, N.~Kulatova, A.~Rastogi, T.~Sibut-Pinote, N.~Swamy, and
  S.~Zanella-B\'{e}guelin, ``Formal verification of smart contracts: Short
  paper,'' in \emph{Proceedings of the 2016 ACM Workshop on Programming
  Languages and Analysis for Security}, ser. PLAS '16.\hskip 1em plus 0.5em
  minus 0.4em\relax New York, NY, USA: Association for Computing Machinery,
  2016, p. 91–96. [Online]. Available:
  \url{https://doi.org/10.1145/2993600.2993611}
\BIBentrySTDinterwordspacing

\bibitem{ICPwhitepaper}
DFINITY, ``The internet computer for geeks,'' (accessed April 15 2024)
  \url{https://eprint.iacr.org/2022/087}, 2022.

\bibitem{NearWhitePaper}
NEAR, ``The near white paper,'' (accessed April 15 2024)
  \url{https://pages.near.org/papers/the-official-near-white-paper/}, 2021.

\bibitem{the_multiversx_team_multiversx_2019}
\BIBentryALTinterwordspacing
T.~M. Team, ``{MultiversX} - {A} {Highly} {Scalable} {Public} {Blockchain} via
  {Adaptive} {State} {Sharding} and {Secure} {Proof} of {Stake},'' Tech. Rep.,
  2019. [Online]. Available:
  \url{https://files.multiversx.com/multiversx-whitepaper.pdf}
\BIBentrySTDinterwordspacing

\bibitem{wood_polkadot_nodate}
G.~Wood, ``\BIBforeignlanguage{en}{{POLKADOT}: {VISION} {FOR} {A}
  {HETEROGENEOUS} {MULTI}-{CHAIN} {FRAMEWORK}},'' Tech. Rep.

\bibitem{hellings_cerberus_2020}
\BIBentryALTinterwordspacing
J.~Hellings, D.~P. Hughes, J.~Primero, and M.~Sadoghi, ``Cerberus:
  {Minimalistic} {Multi}-shard {Byzantine}-resilient {Transaction}
  {Processing},'' Aug. 2020, arXiv:2008.04450 [cs]. [Online]. Available:
  \url{http://arxiv.org/abs/2008.04450}
\BIBentrySTDinterwordspacing

\bibitem{Motoko}
DFINITY, ``Motoko programming language,'' (accessed April 15 2024)
  \url{https://internetcomputer.org/docs/current/motoko/main/motoko}.

\bibitem{ckBTC}
------, ``Chain-key bitcoin (ckbtc),'' (accessed April 15 2024)
  \url{https://internetcomputer.org/docs/current/developer-docs/integrations/bitcoin/ckbtc}.

\bibitem{SatoshiWhitepaper}
S.~Nakamoto, ``Bitcoin: A peer-to-peer electronic cash system,''
  \emph{Decentralized business review}, 2008.

\bibitem{tECDSA}
\BIBentryALTinterwordspacing
J.~Groth and V.~Shoup, ``Design and analysis of a distributed ecdsa signing
  service,'' Cryptology ePrint Archive, Paper 2022/506, 2022,
  \url{https://eprint.iacr.org/2022/506}. [Online]. Available:
  \url{https://eprint.iacr.org/2022/506}
\BIBentrySTDinterwordspacing

\bibitem{UTXOPaper}
S.~Delgado-Segura, C.~P{\'e}rez-Sola, G.~Navarro-Arribas, and
  J.~Herrera-Joancomart{\'\i}, ``Analysis of the bitcoin utxo set,'' in
  \emph{Financial Cryptography and Data Security: FC 2018 International
  Workshops, BITCOIN, VOTING, and WTSC, Nieuwpoort, Cura{\c{c}}ao, March 2,
  2018, Revised Selected Papers 22}.\hskip 1em plus 0.5em minus 0.4em\relax
  Springer, 2019, pp. 78--91.

\bibitem{AlexandrescuLockingPatterns}
A.~Alexandrescu, ``Declarative control flow,'' (accessed April 15 2024)
  \url{https://github.com/CppCon/CppCon2015/blob/18943d6288b1cba54922627b71b7d21d7f1175f1/Presentations/Declarative%20Control%20Flow/Declarative%20Control%20Flow%20-%20Andrei%20Alexandrescu%20-%20CppCon%202015.pdf},
  2015.

\bibitem{MotokoTryFinally}
DFINITY, ``Motoko try / finally construct,'' (accessed Jan 9 2025)
  \url{https://internetcomputer.org/docs/current/motoko/main/reference/language-manual/#try}.

\bibitem{lamport2002specifying}
\BIBentryALTinterwordspacing
L.~Lamport, \emph{Specifying Systems: The TLA+ Language and Tools for Hardware
  and Software Engineers}.\hskip 1em plus 0.5em minus 0.4em\relax
  Addison-Wesley, June 2002. [Online]. Available:
  \url{https://www.microsoft.com/en-us/research/publication/specifying-systems-the-tla-language-and-tools-for-hardware-and-software-engineers/}
\BIBentrySTDinterwordspacing

\bibitem{tlc}
Y.~Yu, P.~Manolios, and L.~Lamport, ``Model checking tla+ specifications,'' in
  \emph{Correct Hardware Design and Verification Methods}, L.~Pierre and
  T.~Kropf, Eds.\hskip 1em plus 0.5em minus 0.4em\relax Berlin, Heidelberg:
  Springer Berlin Heidelberg, 1999, pp. 54--66.

\bibitem{konnov2019tla}
I.~Konnov, J.~Kukovec, and T.-H. Tran, ``Tla+ model checking made symbolic,''
  \emph{Proceedings of the ACM on Programming Languages}, vol.~3, no. OOPSLA,
  pp. 1--30, 2019.

\bibitem{cousineau2012tla}
D.~Cousineau, D.~Doligez, L.~Lamport, S.~Merz, D.~Ricketts, and H.~Vanzetto,
  ``Tla+ proofs,'' in \emph{FM 2012: Formal Methods: 18th International
  Symposium, Paris, France, August 27-31, 2012. Proceedings 18}.\hskip 1em plus
  0.5em minus 0.4em\relax Springer, 2012, pp. 147--154.

\bibitem{lamport2009pluscal}
L.~Lamport, ``The pluscal algorithm language,'' in \emph{International
  Colloquium on Theoretical Aspects of Computing}.\hskip 1em plus 0.5em minus
  0.4em\relax Springer, 2009, pp. 36--60.

\bibitem{ckBTCModel}
O.~Mari\'{c}, ``ckbtc tla+ model,'' (accessed April 17 2025)
  \url{https://github.com/dfinity/formal-models/tree/55f806d396a999a7a244545028f8d21874457f56/tla/ckbtc}.

\bibitem{lu_learning_2008}
\BIBentryALTinterwordspacing
S.~Lu, S.~Park, E.~Seo, and Y.~Zhou, ``Learning from mistakes: a comprehensive
  study on real world concurrency bug characteristics,'' in \emph{Proceedings
  of the 13th international conference on {Architectural} support for
  programming languages and operating systems}, ser. {ASPLOS} {XIII}.\hskip 1em
  plus 0.5em minus 0.4em\relax New York, NY, USA: Association for Computing
  Machinery, Mar. 2008, pp. 329--339. [Online]. Available:
  \url{https://doi.org/10.1145/1346281.1346323}
\BIBentrySTDinterwordspacing

\bibitem{yuan_simple_2014}
\BIBentryALTinterwordspacing
D.~Yuan, Y.~Luo, X.~Zhuang, G.~R. Rodrigues, X.~Zhao, Y.~Zhang, P.~U. Jain, and
  M.~Stumm, ``Simple {Testing} {Can} {Prevent} {Most} {Critical} {Failures}:
  {An} {Analysis} of {Production} {Failures} in {Distributed} {Dataintensive}
  {Systems},'' in \emph{Proceedings of the 11th {Symposium} on {Operating}
  {Systems} {Design} and {Implementation} ({OSDI})}, 2014. [Online]. Available:
  \url{https://www.usenix.org/system/files/conference/osdi14/osdi14-paper-yuan.pdf}
\BIBentrySTDinterwordspacing

\end{thebibliography}

\end{document}